\documentstyle[12pt]{article}

\newcommand{\mysection}{\setcounter{equation}{0}\section}

\begin{document}

\hfill{FERMILAB-Conf-93/117-T}
\vskip 0.5cm
\centerline{\large\bf A Note on  `Hot Spot' Hunting in
Deep-Inelastic Scattering}
\vskip 2cm
\centerline{\sc E. Laenen\footnote{email: eric@fnth010.fnal.gov}
 and E. Levin\footnote{On leave from St. Petersburg Nuclear Physics Institute,
188350 Gatchina, St. Petersburg, Russia\\
e-mail: levin@fnal.fnal.gov, FNAL::LEVIN}}
\vskip 0.8cm
\centerline{\it Fermi National Accelerator Laboratory,}
\centerline{\it P.O. Box 500, MS 106}
\centerline{\it Batavia, Illinois 60510}
\vskip 1cm
\centerline{\it Talk presented \footnote{by E. Levin} at UK Phenomenology
Workshop}
\centerline{\it `HERA - the new frontier for QCD', Durham}
\centerline{\it March, 1993}
\vskip 1cm
\centerline{\bf Abstract}
\vskip 1cm

We consider the  inclusive cross section
for jet production with large transverse momentum in deep-inelastic
scattering. This process has been proposed as a probe
of small-x physics, particularly the measurement of
`hot spots' inside the proton.
We present a numerical calculation of this process,
taking into account a larger phase space.
The theoretical reliability
as well as phenomenological uncertainties of the
calculation are discussed.
\vfill
\newpage

\mysection{Introduction.}
In order to study the main properties of low $x_B$ deep-inelastic
scattering processes, Mueller  suggested in \cite{Mue} an experiment
in which all anticipated new phenomena in this kinematical region
should have a large effect.
His idea was to measure the inclusive production of a gluon jet
with a transverse momentum $k_{j,t}$ very close to the photon
virtuality $Q$ and with a fraction of energy $x_j$ as close to one
as is feasible, so that the ratio $x_B/x_j$ can be small.
In this case,
\begin{itemize}
\item the cross section of the process can be calculated
within the framework of perturbative QCD, if
$k_t^2 \simeq Q^2 >> \Lambda_{\rm QCD}^2$.
\item the dependence of the cross section on $x_B$ is governed
by low $x_B$ gluon emission which can be described by the
BFKL evolution equation \cite{BFKL}. This is in contrast
with the usual GLAP \cite{GLAP} approach, in which the
cross section is described by the simple Born diagram of
Fig.1a. and turns out to be constant with respect to $x_B$.
\item the scale of the shadowing corrections is determined
by the size of the `hot spot' , namely $R \simeq 1/k_{j,t}$
and they are expected to be large (see ref. \cite{GLR}).
\end{itemize}

Numerical estimates for this process have been performed in a
series of papers \cite{BR}-\cite{BL}, but, in our view, the matter has not
been settled yet
(cf. the strong
dependence on an  infrared cut-off in the
BFKL equation with running coupling, for values of $k_t^2$ and $Q^2$ that are
smaller than about 50 ${\rm GeV}^2$ \cite{KMS}-\cite{BL}).

In this paper we reconsider the theoretical formulae for the
cross section and present our numerical estimates in studying
the small-x and  infrared behavior of the one-jet inclusive cross section.
We will not consider any shadowing corrections.

\mysection{The basic formula.}
The correct formula for inclusive one-jet production
in the region of small $x_B$ was given in ref. \cite{Dok}.
In terms of a differential structure function involving the jet
variables $x_j$ and $k_t^2$ (resp. the longitudinal momentum
fraction of the jet and its tranverse momentum squared),
it looks as follows for gluon jet production
\begin{eqnarray}
\frac{dF_2(x_B, Q^2; x_j, k_t^2)}
{d\ln x_j dk_t^2} & = & \frac{3 N_c}{\pi k_t^2} \int
\frac{d^2{\bf k_{1,t}} d^2{\bf k_{2,t}}}{\pi} \cdot
\alpha_S(min\{k_{1,t}^2,k_{2,t}^2,k_{t}^2\}) \nonumber \\
 &\,& \phi_{B}(\frac{x_B}{x_j},k_{1,t}^2,Q^2)
\phi_{G}(x_j,k_{2,t}^2) \cdot \delta^{(2)}({\bf k_t} -
{\bf k_{1,t}} - {\bf k_{2,t}}) \nonumber \\
 & = & \frac{3 N_c}{\pi k_t^2} \int \frac{dk_{1,t}^2 dk_{2,t}^2}
{\sqrt{-\lambda(k_{1,t}^2, k_{2,t}^2, k_t^2)}}\cdot
\alpha_S(min\{k_{1,t}^2,k_{2,t}^2,k_{t}^2\}) \nonumber \\
 &\,& \phi_{B}(\frac{x_B}{x_j},k_{1,t}^2,Q^2)
\phi_{G}(x_j,k_{2,t}^2) ,
\label{eqone}
\end{eqnarray}
where all notation is explained in Fig.1b.,
$\lambda(x,y,z) = x^2+y^2+z^2-2xy-2xz-2yz$,
and $\phi_G$ is the
gluon density, related to the gluon distribution function
$x G(x,Q^2)$ by
\begin{equation}
\frac{d \alpha_S(k^2) x G(x,k^2)}{dk^2} = \phi_G(x,k^2) \alpha_S(k^2).
\label{eqtwo}
\end{equation}
As was shown in ref. \cite{CL}, eq. (\ref{eqone}) can be reduced
 in the double logarithm
approximation (DLA) of perturbative QCD to the
expression of eq.(51) in \cite{Dok}, where the inclusive
production in deep-inelastic scattering was studied in detail
within this approximation. (see also \cite{CL}).
If we integrate in eq. (\ref{eqone}) only the part of phase space where
$k_{2,t}^2 << k_{1,t}^2 $ ($k_{1,t}^2\simeq k_t^2$) (`small' phase space)
we can rewrite eq. (\ref{eqone}) in the form:
\begin{eqnarray}
\frac{dF_2}{d\ln x_j dk_t^2} & = & \frac{3 N_c}{\pi k_t^2} \int
dk_{2,t}^2 \cdot
\alpha_S(k_{2,t}^2)
 \phi_{B}(\frac{x_B}{x_j},k_t^2,Q^2)
\phi_{G}(x_j,k_{2,t}^2) \nonumber \\
& = & \frac{3 N_c}{\pi k_t^2}\alpha_S(k_t^2)
\phi_{B}(\frac{x_B}{x_j},k_t^2,Q^2)
x_j G(x_j,k_t^2). \label{eqthree}
\end{eqnarray}
It is this equation that was used in all previous numerical estimates
of the inclusive jet production \cite{BR}-\cite{BL}. Here we calculate the
differential structure function without any restriction on the
region of integration (`large' phase space).
In comparing the two we will find an enhancement of the
cross section due to the larger phase space of about 80 \%.

\mysection{Calculational procedure.}
The functions $\phi$ in eq. (\ref{eqone}) are solutions of the BFKL-equation
\cite{BFKL},
\begin{equation}
\frac{\partial\phi(x,k^2)}{\partial\ln(1/x)} = \frac{3\alpha_S}{\pi}
\int_{k_0^2}^{\infty} dk'^2 \Big\{ \frac{\phi(x,k'^2)-\phi(x,k^2)k^2/k'^2}
{|k'^2 - k^2|} + \frac{k^2}{k'^2}\frac{\phi(x,k^2)}{(4k'^4+k^4)^{(1/2)}}
\Big\}. \label{eqfour}
\end{equation}
We now discuss aspects of our procedure of solving this
equation.

We chose different initial conditions
for the functions $\phi_{B}(\frac{x_B}{x_j},k_t^2,Q^2)$ and
$\phi_{G}(x_j,k_t^2)$
in (\ref{eqone}). For $\phi_{B}(\frac{x_B}{x_j},k_t^2,Q^2)$ we
used the same initial condition as in ref. \cite{KMS}, namely
at $z_0 = x_B/x_j = 10^{-1}$
\begin{equation}
\phi_B(z_0,k_t^2,q^2) = \frac{F_0(z_0,k_t^2,Q^2)}{k^2} \simeq
\frac{F_0(k^2,Q^2)}{k^2},
\end{equation}
where the function $F_0$, related to the quark box diagram, was calculated
in refs. \cite{BR}-\cite{KMS}.
For $\phi_{G}(x_j,k_t^2)$ we have to reconstruct the initial
condition from experimental data. However, in order to solve the
BFKL-equation we need to know the behavior of $\phi$ at any value
of $k_t^2$ for some fixed $x$, even at $k_t^2 \rightarrow 0$.
To accomplish this, we used the following procedure to describe
the low $k_t^2$ behavior of $x_j G(x_j,k_t^2)$:
\begin{equation}
x_j G(x_j,k^2) \rightarrow \frac{k^2}{k^2 + q_0^2}
{\tilde x}_j G({\tilde x}_j,{\tilde k}^2)
\label{eqsix}
\end{equation}
where ${\tilde k}^2 = k^2 + q_0^2$, ${\tilde x} = x/(x+k^2/{\tilde k}^2(1-x))$.
We used the mapping (\ref{eqsix}) because (1) this parametrization ensures $x_j
G(x_j,k^2)
\simeq k^2$ as $k^2\rightarrow 0$ and such a behavior is the direct
consequence of the gauge invariance of QCD, and (2) it works for the case
of $F_2$ (see refs. \cite{Bod}-\cite{ALLM}).

For the function ${\tilde x}_j G({\tilde x}_j,{\tilde k}^2)$ we
used a fit \footnote{We are grateful to
J. Botts for making this fit, and discussions relating to it.}
to the  data set from the CTEQ collaboration \cite{CTEQ}
down to $Q^2$ values of 1 GeV$^2$.
The initial condition for $\phi_G$ was then constructed according
to (\ref{eqtwo}) (we used a fixed coupling), at a value
$x=10^{-2}$, where, as in \cite{KMS}, instead of $xG(x,k^2)$ we used
the effective density $xG(x,k^2) + \frac{4}{9}\sum_{f=1}^4 x
[q_f(x,k^2) + \bar{q}_f(x,k^2)]$.

We solved\footnote{We are grateful to J. Kwiecinski for sending us
his program to solve the BFKL equation numerically.}
 the BFKL equation (\ref{eqfour}) with a fixed coupling
$\alpha_S = \alpha_S(Q^2)$ because this equation
only sums the leading logs $(\alpha_S \ln(1/x))^n$, and not the
subleading ones. The latter remains an unsolved problem.
However, we made a rough estimate of how important the
corrections from a running $\alpha_S$ could be by
calculating an average $\alpha_S$
\begin{equation}
<\alpha_S>_i = \frac{\int dk^2 \alpha_S(k^2) \phi_i(x,k^2)}
{\int dk^2 \phi_i(x,k^2)} \quad,\quad i=B,G
\end{equation}
The dependence of $<\alpha_S>_i$ on $x_B$ showed that we
could use $\alpha_S= \alpha_S(Q^2)$ as a good first approximation.
For example, for $k^2=Q^2 =10$ GeV$^2$, we found
$\alpha_S(Q^2) \simeq 0.2$, and $<\alpha_S>_i \simeq 0.15$
for $x < 10^{-2}$ for both $i = B,G$.

A much more detailed study of this problem can be found in
\cite{BL}.

The last point we discuss is the infrared cut-off $k_0^2$
in eq. (\ref{eqfour}). In principle eq. (\ref{eqfour}) is infrared
stable and one can take $k_0^2=0$. However, in order to further
investigate the dependence of the solution of (\ref{eqfour})
on small momenta we preferred to introduce a cut-off $k_0^2$
and see how much the answer depends on its value.

\mysection{Results and conclusions.}

The results are shown in Figs.2a and 2b. for
the values $k^2 = Q^2 = 10$ GeV$^2$, $x_j = 10^{-2}$. The first observation
we make from Fig.2a is that the answer for (\ref{eqone}) when one
includes the correct (`large') phase space increases
the results of previous calculations (small phase space)
by about 80 \%. Furthermore, the dependence on $q_0^2$ is visible,
but does not compensate for the difference between small
and large phase space. A similar conclusion we found to hold for
the dependence on the infrared cut-off $k_0^2$.

The most discouraging result is shown in Fig.2b. Here
we plot the normalized differential structure function
\begin{equation}
	R_2 = \frac{1}{F_2(x_B,Q^2)}\frac{dF_2(x_B,Q^2; x_j, k^2)}
{d\ln x_j dk^2}
\end{equation}
where we took $F_2(x,Q^2) = \int dk^2 \phi_{B}(\frac{x_B}{x_j},k^2,Q^2)
\phi_{G}(x_j,k^2)$.
One notes that this ratio seems to be
independent of $x_B$. This would seem to indicate that,
within the approximations made, this special environment
does not seem to be much better for measuring small-x
effects than a direct measurement of $F_2$. This feature
persisted when we took $F_2(x,Q^2)$ constructed from the
MRS ${\rm D}\_'$ distributions \cite{MRS}.

Figure Captions
\begin{description}
\item[Fig.1a.]
Born diagram for single gluon jet production in deep-inelastic scattering.
\item[Fig.1b.]
Single-jet production in deep-inelastic scattering in
the large phase space case.
\item[Fig.2a.]
Plot of $\frac{dF_2}{d\ln x_j dk^2}$ vs. $x_B$. The value of
$x_j$ is 0.01 and we took $k^2= Q^2 = 10$ GeV$^2$. The solid line
corresponds to the small phase space case, the three
remaining to the large phase space case with three different
values of $q_0^2$, namely $q_0^2 = 1$ GeV$^2$ (long-dashed),
$q_0^2 = 2$ GeV$^2$ (short-dashed) and $q_0^2 = 4$ GeV$^2$ (dotted).
\item[Fig.2b.]
The ratio $R_2$ (4.1) for three values of $q_0^2$ in
the large phase space case. The notation is the same
as in Fig.2a.
\end{description}

\end{document}